\documentclass{article}

\usepackage{arxiv}

\usepackage[utf8]{inputenc} 
\usepackage[T1]{fontenc}    
\usepackage{hyperref}       
\usepackage{url}            
\usepackage{booktabs}       
\usepackage{amsfonts}       
\usepackage{nicefrac}       
\usepackage{microtype}      
\usepackage{lipsum}
\usepackage{amsmath,amssymb,amsfonts}
\usepackage{color}
\usepackage{algorithmic}
\usepackage{graphicx}
\usepackage{textcomp}
\usepackage{comment,color}
\usepackage{enumitem}
\usepackage{epsfig}
\usepackage{caption}
\usepackage{verbatim,bm,multicol,blindtext}
\usepackage{epstopdf,array,mathtools,url,subfigure}

\title{PINNtomo: Seismic tomography using physics-informed neural networks}

\author{
  Umair bin Waheed\\
  Department of Geosciences\\
  King Fahd University of Petroleum and Minerals\\
  Dhahran 31261, Saudi Arabia. \\
  \texttt{umair.waheed@kfupm.edu.sa} \\
   \And
 Tariq Alkhalifah \\
  Physical Sciences and Engineering Division\\
  King Abdullah University of Science and Technology\\
  Thuwal 23955, Saudi Arabia.
   \And
  Ehsan Haghighat\\
  Department of Civil Engineering \\
  Massachusetts Institute of Technology\\
  MA 02139, USA.
   \And
 Chao Song \\
  Physical Sciences and Engineering Division\\
  King Abdullah University of Science and Technology\\
  Thuwal 23955, Saudi Arabia.
   \And
 Jean Virieux\\
  ISTERRE\\
  Universit\'{e} Grenoble Alpes\\
  Saint-Martin-d’Heres 38400, France.
}

\begin{document}
\maketitle

\begin{abstract}
Seismic traveltime tomography using transmission data is widely used to image the Earth's interior from global to local scales. In seismic imaging, it is used to obtain velocity models for subsequent depth-migration or full-waveform inversion. In addition, cross-hole tomography has been successfully applied for a variety of applications, including mineral exploration, reservoir monitoring, and CO$_2$ injection and sequestration. Conventional tomography techniques suffer from a number of limitations, including the use of a smoothing regularizer that is agnostic to the physics of wave propagation. Here, we propose a novel tomography method to address these challenges using developments in the field of scientific machine learning. Using seismic traveltimes observed at seismic stations covering part of the computational model, we train neural networks to approximate the traveltime factor and the velocity fields, subject to the physics-informed regularizer formed by the factored eikonal equation. This allows us to better compensate for the ill-posedness of the tomography problem compared to conventional methods and results in a number of other attractive features, including computational efficiency. We show the efficacy of the proposed method and its capabilities through synthetic tests for surface seismic and cross-hole geometries. Contrary to conventional techniques, we find the performance of the proposed method to be agnostic to the choice of the initial velocity model.
\end{abstract}

\keywords{Tomography \and Inversion \and Traveltimes \and Neural networks \and Machine learning}

\section{Introduction}

Seismic tomography has been used over the years as a pre-eminent tool for subsurface model building at various scales ranging from global and regional scales in earthquake seismology~\cite{rawlinson2010seismic} to local scales in exploration seismology~\cite{stewart1991exploration}. Building a velocity macro-model is a crucial step for the success of depth migration~\cite{etgen2009overview} and full-waveform inversion~\cite{virieux2009overview} for high-resolution imaging of the Earth's crust. First arrival traveltime tomography based on refraction data or diving waves has been successfully used to build such initial models~\cite{taillandier2009first}. In particular, the method is attractive for land seismic data processing with surface acquisition, since it is often difficult or even impossible to identify reflections in the data. Moreover, cross-hole seismic tomography, which uses direct arrival times, has been around for more than three decades in oil/gas~\cite{bregman1989crosshole} and mineral exploration~\cite{wong2000crosshole}. It has also been successful for applications in void and tunnel detection~\cite{rechtien1995tunnel}, reservoir characterization~\cite{yamamoto2001imaging}, fracture detection~\cite{maurer1997potential}, hydrological parameter estimation~\cite{hyndman2000inferring}, geotechnical site investigations~\cite{ng2019potential}, and time-lapse studies related to carbon capture and sequestration~\cite{ajo2013high}. 

Seismic tomography is typically solved as an inverse problem that minimizes the misfit between a set of observed arrival times on the receivers and those synthetically generated using an estimate of the velocity model. Minimization of this misfit function requires a nonlinear optimization procedure. However, the conventional ray tomography approach linearizes the tomography operator, which requires the computation of the Fr\'{e}chet derivatives. Then the linearized tomography operator is inverted iteratively. For modern seismic surveys, the requirement of explicitly computing the Fr\'{e}chet derivatives is challenging to handle in terms of the computation cost and memory requirements. This gave way to the adjoint-state method~\cite{sei1994gradient,leung2006adjoint} that formulates tomographic inversion as a nonlinear optimization process by directly computing the gradient of the misfit function. 

Nevertheless, these conventional methods still suffer from a number of limitations. Usually, they use some form of smoothing regularization to compensate for the ill-posedness of the problem. This ends up limiting the resolution of the inverted velocity model. In addition, these methods typically need an initial model with some general background features of the Earth represented, like a constant depth gradient. The choice of the initial model may affect the final solution and is usually not obvious prior to inversion. Furthermore, for models with irregular topography, considerable grid and algorithmic adaptions are needed to account for the free-surface topography~\cite{ma2015topography}. 

Therefore, in this work, we propose a novel algorithm for the seismic tomography problem based on developments in the field of scientific machine learning. In particular, we use the emerging paradigm of physics-informed neural networks (PINNs) that overcomes the limitation of deep learning associated with sparse data by incorporating the governing partial differential equation (PDE) into the neural network's loss function. PINNs have already demonstrated success in solving a number of forward and inverse problems in other scientific disciplines~\cite{raissi2020hidden,sahli2020physics}. Recently, PINNs have also shown remarkable success in overcoming limitations associated with conventional techniques in modeling seismic traveltimes~\cite{bin2020eikonal,smith2020eikonet} and wavefields~\cite{moseley2020solving,song2021solving}. 

Here, we develop a PINN-based tomography (PINNtomo) algorithm to invert for the velocity model. Given traveltimes at seismic stations covering part of the computational domain, we use neural networks to approximate the traveltime factor and the velocity fields, subject to the physics-informed regularizer based on the factored eikonal equation. Doing so allows us to better compensate for the poorly determined aspects of the velocity model compared to conventional physics-agnostic smoothing regularizers. Also, we find the performance of the method to be independent of the initial velocity model. Moreover, since the method is mesh-free, it is easily adaptable to models with irregular topography without modifications. Additional advantages of the method include ease of deployment across a variety of platforms (CPUs, GPUs) and architectures (desktops, clusters) without any modification.

Through tests on realistic surface seismic and cross-hole geometries, we demonstrate the efficacy of the proposed algorithm in solving the tomography problem. This is done by obtaining a velocity model that produces traveltimes matching those observed at seismic stations while honoring the physics of wave propagation by minimizing the residual of the eikonal equation at selected grid points in the computational domain. 

\section{Theory}

In an isotropic medium, the eikonal equation relates the gradient of the traveltime surfaces to the velocity of the wavefront through the relation:
\begin{equation}
\begin{aligned}
    |\nabla T (\mathbf{x})|^2 & = \frac{1}{v^2(\mathbf{x})}, \qquad \forall \, \mathbf{x}\, \in \, \Omega, \\
   T(\mathbf{x}_s) & = 0,
\label{eq:eikonal}
\end{aligned}
\end{equation}
where $\Omega$ is a domain in $\mathbb{R}^d$ with $d$ as the space dimension, $T(\mathbf{x})$ is the traveltime or Euclidean distance to any point $\mathbf{x}$ from the point-source $\mathbf{x}_s$, $v(\mathbf{x})$ is the velocity defined on $\Omega$, and $\nabla$ denotes the gradient operator.

Since equation~\eqref{eq:eikonal} contains singularity at the point-source location, traveltime modeling studies~\cite{bin2020eikonal,smith2020eikonet} have shown that a factored form of the eikonal equation is easier to train using PINNs. Therefore, we factorize the traveltime $T(\mathbf{x})$ into two multiplicative functions~\cite{fomel2009fast}, i.e.,
\begin{equation}
    T(\mathbf{x}) = T_0(\mathbf{x}) \, \tau(\mathbf{x}),
    \label{eq:factorization}
\end{equation}
where $T_0(\mathbf{x})$ is the known function which is computed analytically, leaving $\tau(\mathbf{x})$ as the unknown traveltime factor. By substituting the above in equation~\eqref{eq:eikonal}, we get the factored eikonal equation:
\begin{equation}
\begin{aligned}
    T_0^2\,|\nabla \tau|^2 + \tau^2\,|\nabla T_0|^2 & + 2 \, T_0 \, \tau \, (\nabla T_0 . \nabla \tau)  = \frac{1}{v^2(\mathbf{x})},\\
   \tau(\mathbf{x}_s) & = 1.
\label{eq:fac_eikonal}
\end{aligned}
\end{equation}
The known traveltime $T_0$ is computed analytically using the expression:
\begin{equation}
    T_0(\mathbf{x}) = \frac{|\mathbf{x} - \mathbf{x}_s|}{v(\mathbf{x}_s)},
    \label{eq:knownsol}
\end{equation}
where $v(\mathbf{x}_s)$ is the velocity at the source location. This ensures that $T_0$ captures the point-source singularity leaving $\tau$ as a smooth function in the source neighborhood.

We can re-write the factored eikonal equation in its residual form as:
\begin{equation}
    L(\mathbf{x}) : T_0^2\,|\nabla \tau|^2 + \tau^2\,|\nabla T_0|^2 + 2 \, T_0 \, \tau \, (\nabla T_0 . \nabla \tau)  - \frac{1}{v^2(\mathbf{x})} = 0.
\label{eq:residual}
\end{equation}

To invert for the unknown velocity model, we consider two multilayer neural networks -- one to approximate the unknown traveltime factor for an arbitrary source location $\mathbf{x}_s$, $\tau(\mathbf{x}_s,\mathbf{x})$, and the other for the velocity, $v(\mathbf{x})$, i.e.,
\begin{equation}
\begin{aligned}
    \tau(\mathbf{x}_s,\mathbf{x}) \approx \hat{\tau}(\mathbf{x}_s,\mathbf{x}) & = \mathcal{N}_\tau(\mathbf{x}_s, \mathbf{x}; \boldsymbol{\theta}_\tau), \\
    v(\mathbf{x}) \approx \hat{v}(\mathbf{x}) & = \mathcal{N}_v(\mathbf{x}; \boldsymbol{\theta}_v),
\end{aligned}
\end{equation}
where $\mathcal{N}_\tau$ and $\mathcal{N}_v$ are the neural networks with trainable parameters $\boldsymbol{\theta}_\tau$ and $\boldsymbol{\theta}_v$, respectively. Since traveltime data from multiple sources are needed to obtain a reliable velocity model, the traveltime factor network also takes shot locations as input, in addition to the spatial coordinates. On the contrary, the velocity model only takes spatial coordinates as input since it does not vary with the source locations.

Since both the traveltime factor and velocity are strictly positive quantities, we pass the output of the networks through a sigmoid function, $\sigma()$, and multiply these by scaling coefficients, i.e.,
\begin{equation}
\begin{aligned}
    \hat{\tau}(\mathbf{x}_s,\mathbf{x}) & = \sigma \left(\mathcal{N}_\tau(\mathbf{x}_s, \mathbf{x}; \boldsymbol{\theta}_\tau)\right) \, \tau_{\mathrm{peak}}, \\
    \hat{v}(\mathbf{x}) & = \sigma \left(\mathcal{N}_v(\mathbf{x}; \boldsymbol{\theta}_v)\right) \, v_{\mathrm{peak}},
\end{aligned}
\end{equation}
where $\tau_{\mathrm{peak}}$ and $v_{\mathrm{peak}}$ are the peak values that can be obtained from the traveltime factor network and the velocity network, respectively. These scaling factors should be chosen such that they are larger than the expected maximum values to avoid clipping of the output.

\begin{figure}[ht!]
\begin{center}
\includegraphics[width=0.7\textwidth]{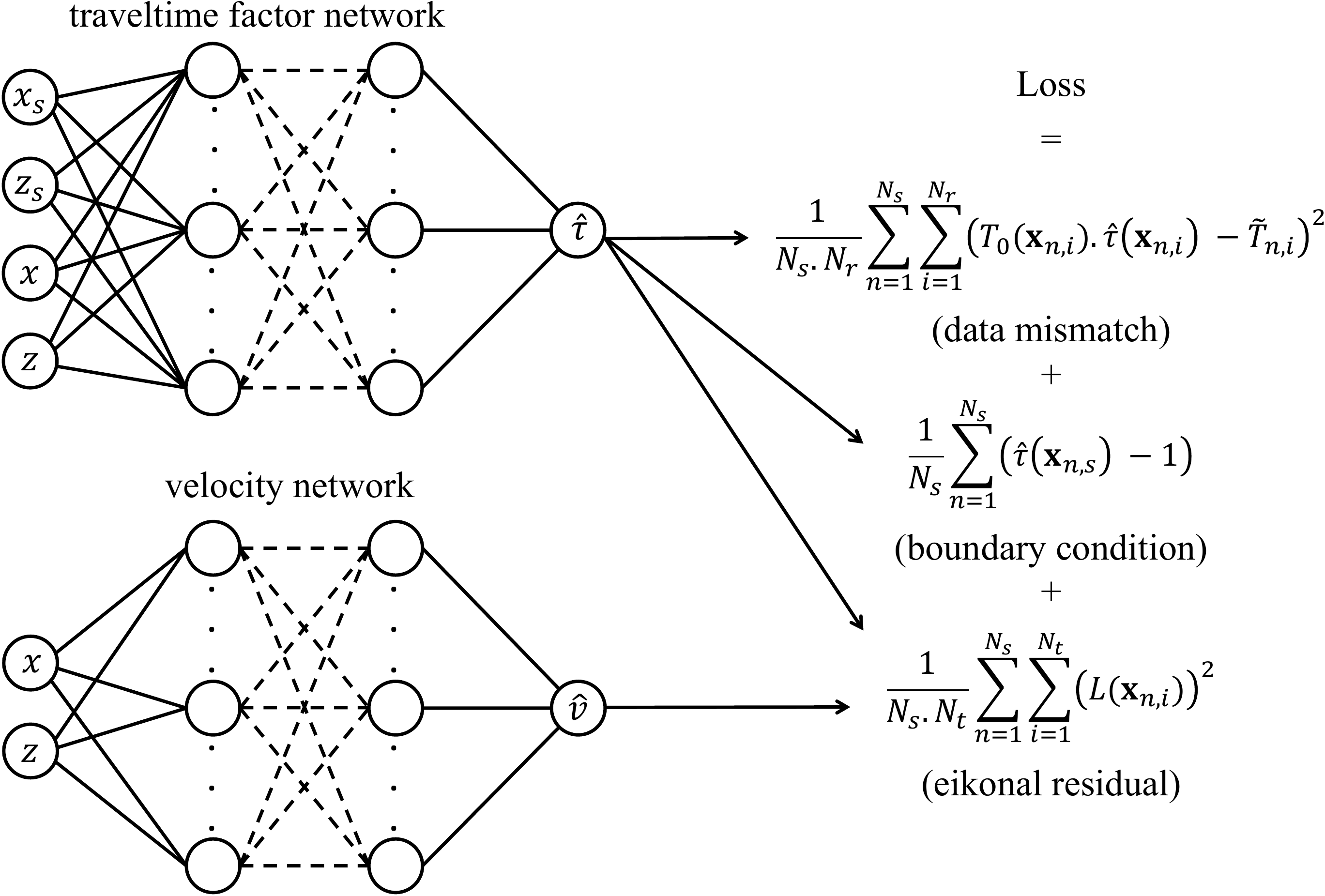}
\end{center}
\caption{
Illustration of the PINNtomo algorithm: We use two neural networks to approximate the traveltime factor $\hat{\tau}$ and the velocity $\hat{v}$. The loss function used to train these networks minimizes the traveltime mismatch at the receivers (first term), honors the boundary condition (second term), and minimizes the residual of the eikonal equation (third term) at selected training points within the computational domain. These quantities are minimized for all source locations $\mathbf{x}_s = (x_s,z_s)$.
}%
\label{fig:work_flow}
\end{figure}

Finally, we use a single loss function to train both networks simultaneously. The loss function is given as:
\begin{equation}
\begin{aligned}
\mathfrak{J}(\boldsymbol{\theta}_\tau &,\boldsymbol{\theta}_v) =
\frac{1}{N_s\,N_r}\sum_{n = 1}^{N_s} \sum_{i = 1}^{N_r} \left(T_0(\mathbf{x}_{n,i})\,\tau(\mathbf{x}_{n,i}) - \tilde{T}_{n,i} \right)^2 \\
+ & \frac{1}{N_s}\sum_{n = 1}^{N_s} \left(\hat{\tau}(\mathbf{x}_{n,s}) - 1\right)^2 + \frac{1}{N_s\,N_t}\sum_{n = 1}^{N_s} \sum_{i = 1}^{N_t} \left(L(\mathbf{x}_{n,i}) \right)^2 ,
\end{aligned}
\label{eq:loss_mse}
\end{equation}
where $N_s$ denotes the total number of sources, $N_r$ is the total number of receivers, and $N_t$ is the number of training (collocation) points from the computational domain. The first term minimizes the misfit between traveltimes predicted by the neural network and the observed traveltimes, $\tilde{T}_{n,i}$, over all sources and receivers. The second term enforces the boundary condition for all source positions, while the third term ensures that the outputs of the neural networks minimize the residual of the eikonal equation over all sources and training points. Figure~\ref{fig:work_flow} summarizes the proposed loss functions and the neural networks used. 

The network parameters $\boldsymbol{\theta}_\tau$ and $\boldsymbol{\theta}_v$ are then identified by solving the following minimization problem:
\begin{equation}
\arg\min_{\boldsymbol{\theta}_\tau ,\boldsymbol{\theta}_v} \; \mathfrak{J}(\boldsymbol{\theta}_\tau ,\boldsymbol{\theta}_v).
\label{eq:optimization}
\end{equation}

\section{Numerical Tests}

In this section, we test the proposed tomography workflow on cross-hole and surface acquisition geometries. For both tests, we use neural networks containing 10 hidden layers. For the traveltime factor network, each layer contains 20 neurons, whereas, for the velocity network, each layer contains 10 neurons. We use a locally adaptive exponential linear unit (l-ELU) as the activation function for the hidden layers. Locally adaptive activation functions have been shown to achieve superior performance and convergence speed over base methods~\cite{jagtap2020locally}. We train the networks first using the Adam optimizer with a batch size of 1024 for 500 epochs and then using the L-BFGS-B optimizer until convergence. These hyper-parameters are chosen based on some initial tests and kept fixed throughout the study to avoid the need for tuning. The PINN framework is implemented using the \texttt{SciANN} package~\cite{haghighat2021sciann} -- a high level $\texttt{Tensorflow}$ wrapper for scientific computations.

\subsection{Example 1: Cross-hole tomography}

\begin{figure}
  \centering
  \subfigure[]{\includegraphics[width=0.4\textwidth]{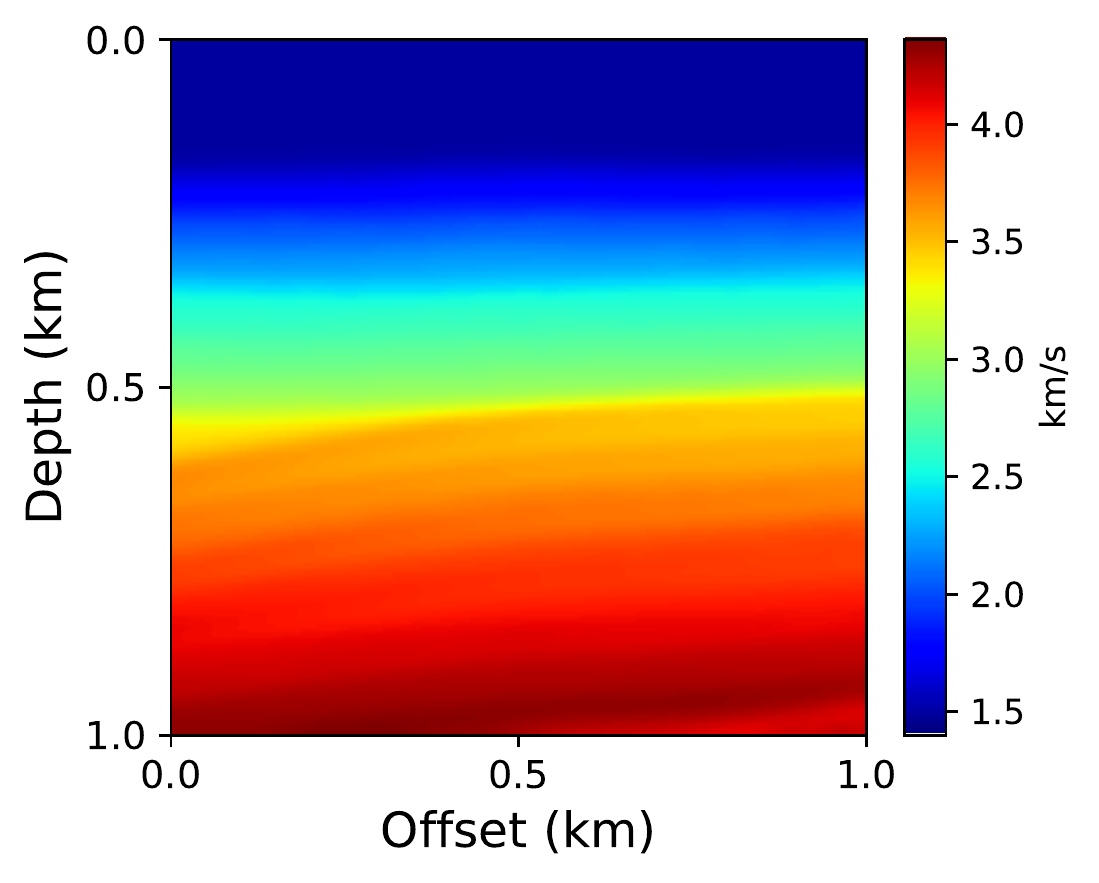}\label{fig:seam_model/v_true}}
  \subfigure[]{\includegraphics[width=0.4\textwidth]{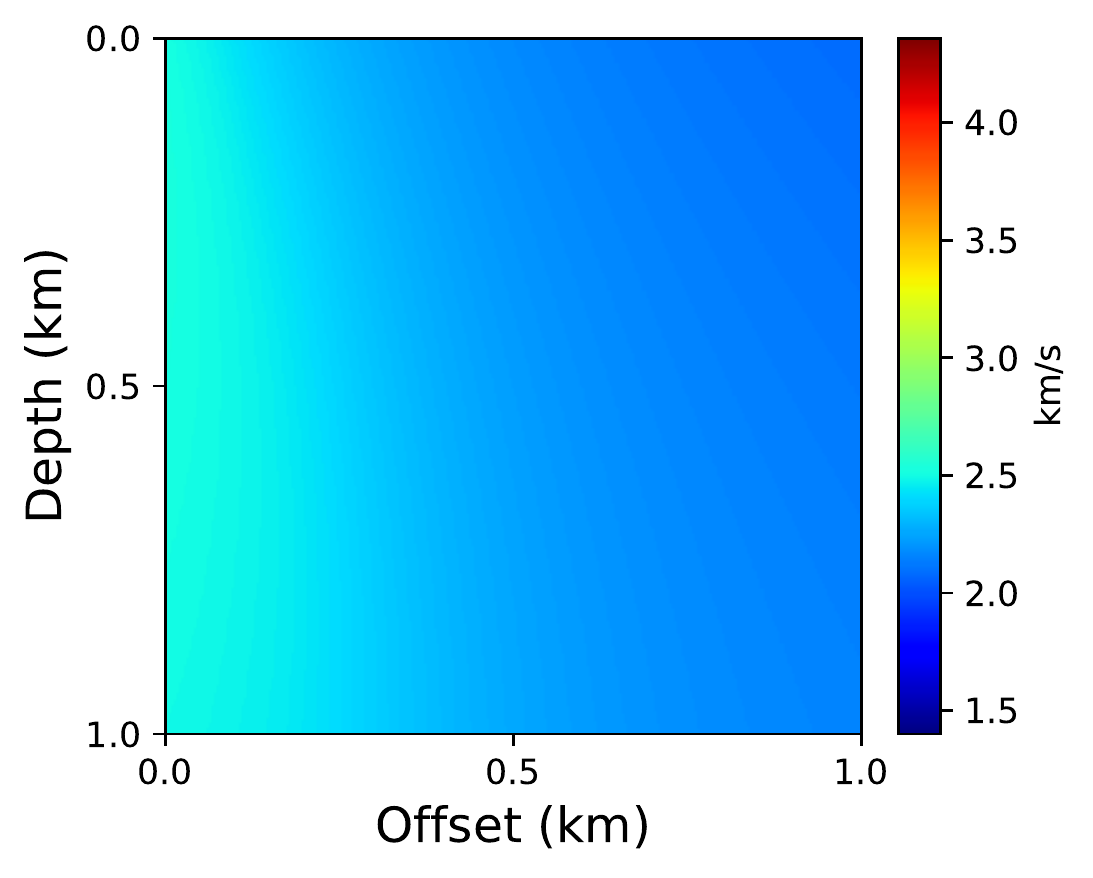}\label{fig:seam_model/vel_ini}}
  \caption{The true (a) and initial (b) velocity models used for the cross-hole tomography test. For this test, we use 11 equispaced sources on the left boundary ($x = 0$~km) of the model and 51 equispaced receivers on the right boundary ($x = 1$~km) of the model.}
  \label{fig:seam_model}
\end{figure}

First, we apply the tomography workflow on a $1\times1$~km$^2$ computational domain with the velocity model shown in Figure~\ref{fig:seam_model/v_true}. We consider 11 sources on the left boundary of the model ($x = 0$~km) uniformly spaced with an interval of 100~m, and 51 receivers on the right boundary of the model ($x = 1$~km) with a uniform spacing of 20~m. The observed traveltimes are computed through a first-order factored eikonal solver using the fast sweeping method~\cite{fomel2009fast}. The initial velocity model for this test is shown in Figure~\ref{fig:seam_model/vel_ini}. It is worth noting that, contrary to conventional tomography methods, here, the initial velocity model is automatically selected based on the initialization of the parameters of the velocity network. While the choice of the initial velocity model is critical to the success of ray tomography, we find our formulation to be agnostic to this choice. For illustration, the initial velocity model is plotted using the same color scale as the true model.

\begin{figure}[ht!]
\begin{center}
\includegraphics[width=0.5\textwidth]{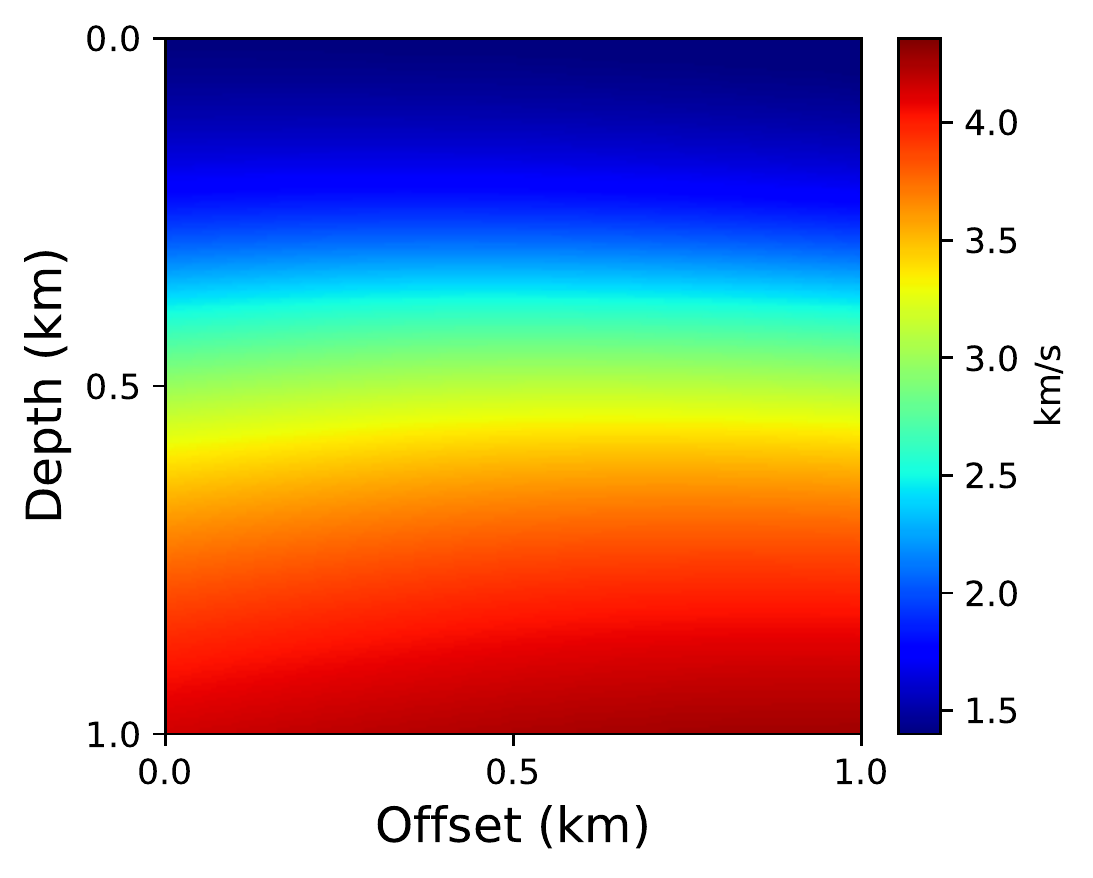}
\end{center}
\caption{
The inverted velocity model for the cross-hole tomography test indicating reliable reconstruction of the long-wavelength features of the true velocity model.
}%
\label{fig:seam_vel_inv}
\end{figure}

\begin{figure}[ht!]
  \centering
  \subfigure[]{\includegraphics[width=0.25\textwidth]{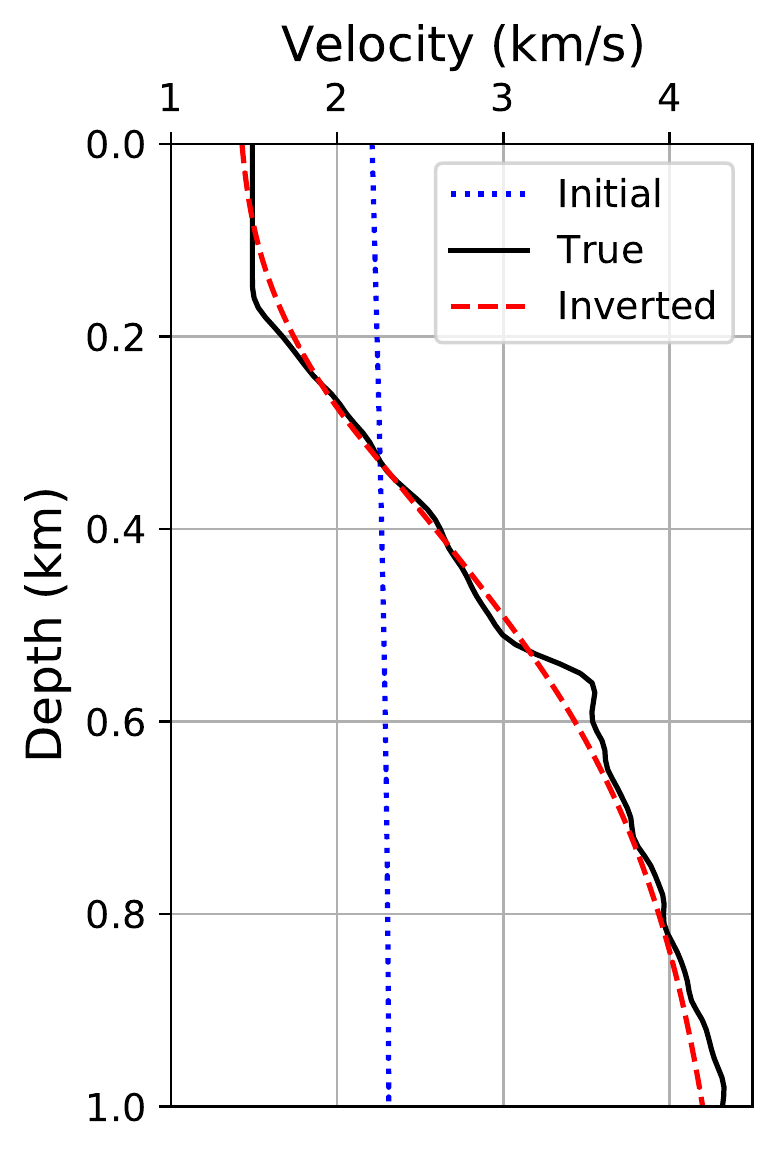}\label{fig:seam_model/vel_trace1}}
  \subfigure[]{\includegraphics[width=0.25\textwidth]{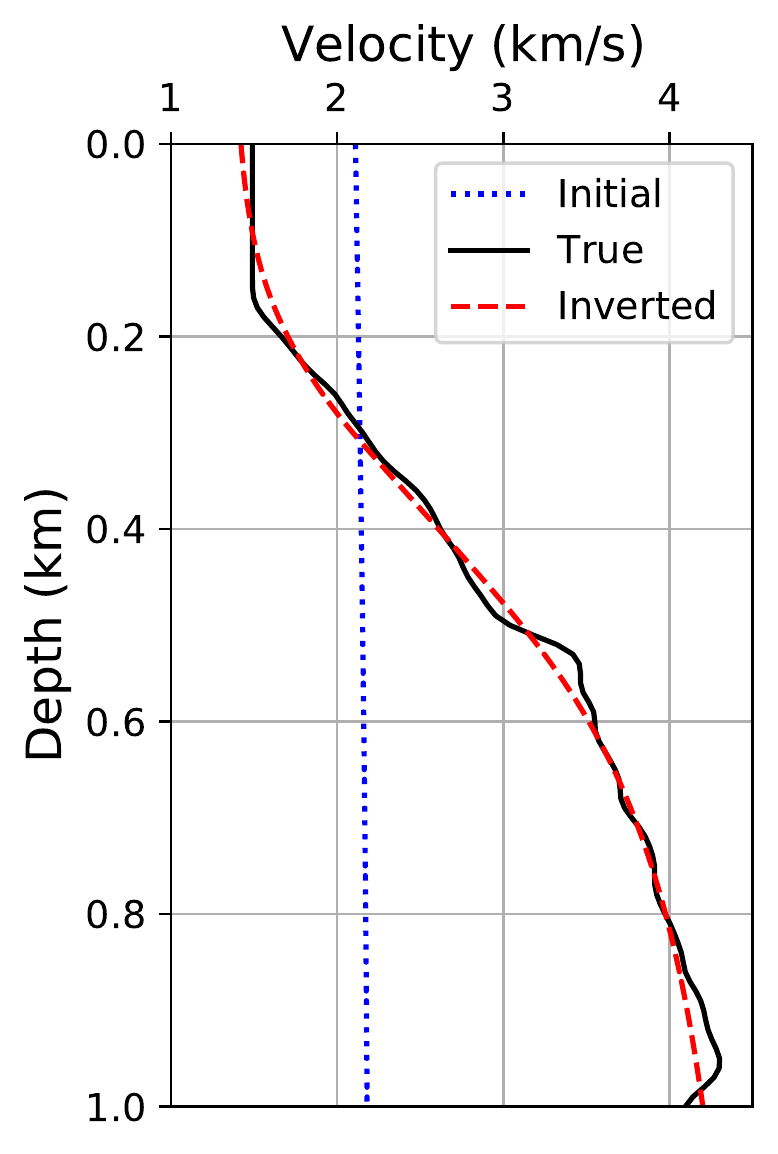}\label{fig:seam_model/vel_trace2}}
  \caption{A comparison of the velocity profiles at $x~=~0.4$~km (a) and $x~=~0.8$~km (b) between the initial (dotted blue), true (solid black), and inverted (dashed red) velocity models.}
  \label{fig:seam_traces}
\end{figure}

\begin{figure}[ht!]
\begin{center}
\includegraphics[width=0.3\textwidth]{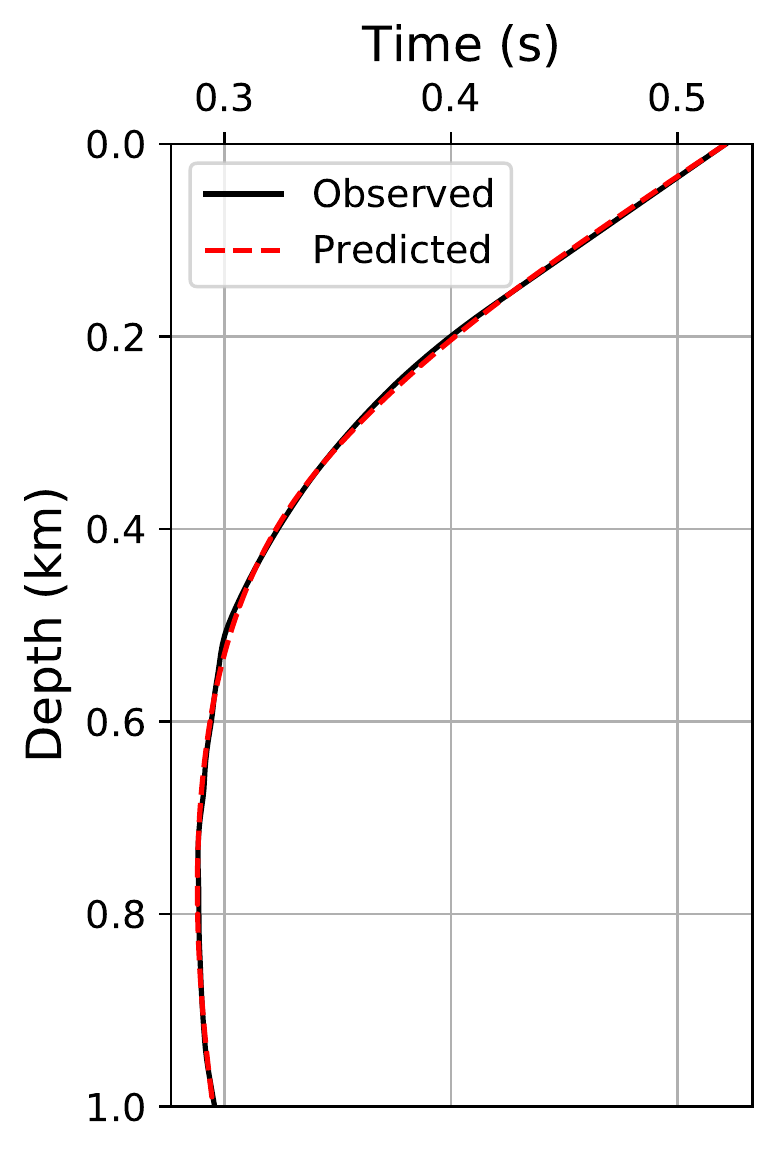}
\end{center}
\caption{
 A comparison between the observed traveltimes (solid black) and those predicted using the neural network (dashed red) at the receivers located on the right boundary of the model for the source at $(x_s,z_s)=(0~\mathrm{km},0.5~\mathrm{km})$.
}%
\label{fig:seam_datafit}
\end{figure}

Figure~\ref{fig:seam_vel_inv} shows the inverted velocity model, indicating that the long-wavelength features of the true model have been well-recovered, as expected from traveltime tomography. Figure~\ref{fig:seam_traces} compares the velocity profiles at $x~=~0.4$~km and $x~=~0.8$~km between the initial, true, and inverted velocities. We observe that the initial velocity model evolved significantly to match the long-wavelength trend of the true velocity model. Finally, in Figure~\ref{fig:seam_datafit} we show the final fit between the observed and the predicted traveltimes at all receivers for the source at $(x_s,z_s)=(0~\mathrm{km},0.5~\mathrm{km})$, indicating good agreement between the two.

\subsection{Example 2: Surface tomography}

\begin{figure}
  \centering
  \subfigure[]{\includegraphics[width=0.85\textwidth]{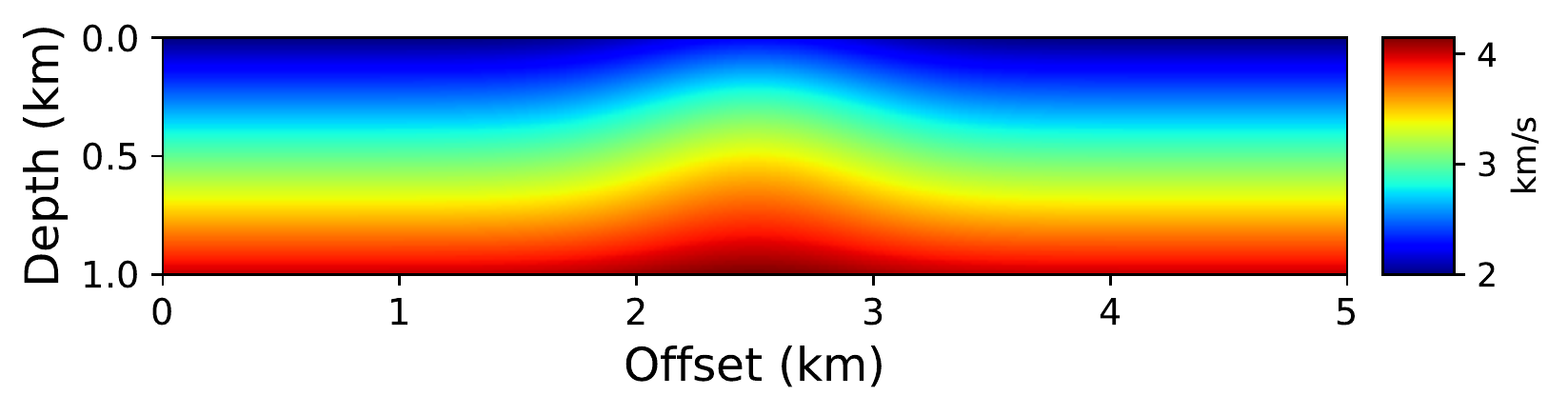}\label{fig:lens_model/v_true}}
  \subfigure[]{\includegraphics[width=0.85\textwidth]{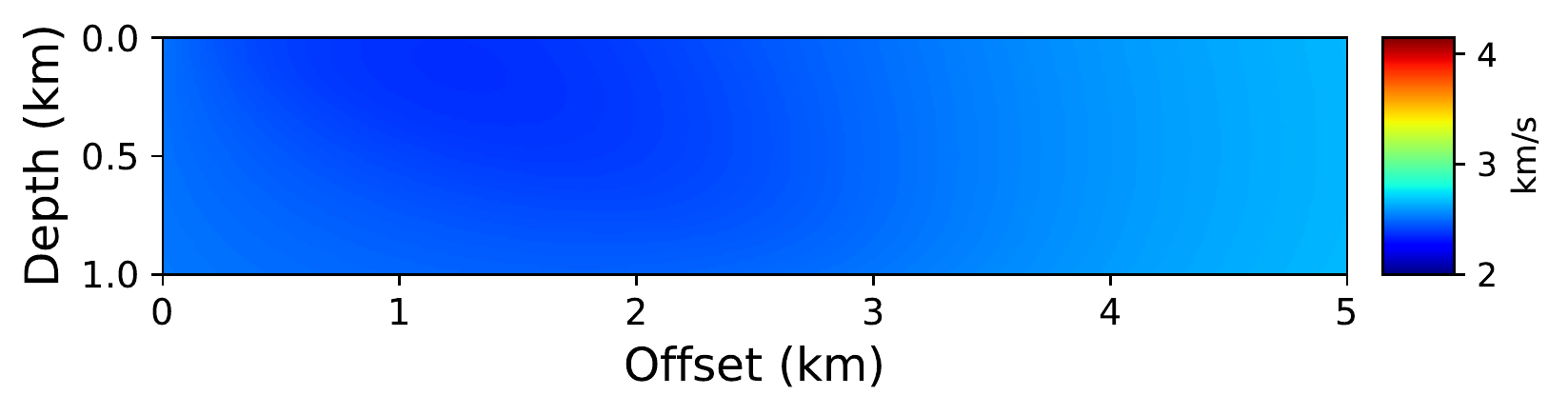}\label{fig:lens_model/vel_ini}}
  \caption{The true (a) and initial (b) velocity models used for the surface tomography test. For this test, we use 51 equispaced sources and 126 equispaced receivers on the surface ($z = 0$~km) of the model.}
  \label{fig:lens_model}
\end{figure}

Next, we test the performance of PINNtomo using a surface acquisition geometry. We consider a $1\times5$~km$^2$ model with the velocity distribution as shown in Figure~\ref{fig:lens_model/v_true}. We consider 51 sources with a uniform spacing of 100~m and 126 receivers with a uniform spacing of 40~m, both on the surface of the model~($z = 0$~km). Like earlier, the observed traveltimes are computed through a first-order factored eikonal solver using the fast sweeping method. Again, the initial velocity is a consequence of the random initialization of the velocity network's parameters. For illustration, the initial velocity model is shown in Figure~\ref{fig:lens_model/vel_ini} using the same color scale as the true model.

\begin{figure}[ht!]
\begin{center}
\includegraphics[width=0.85\textwidth]{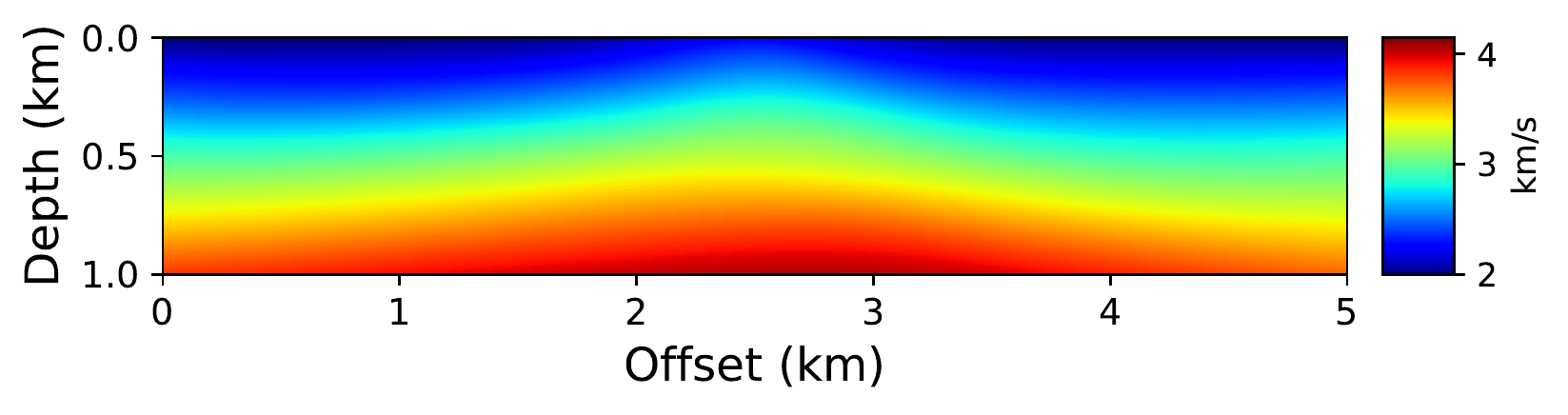}
\end{center}
\caption{
The inverted velocity model for the surface tomography test indicating reliable reconstruction of the long-wavelength features of the true model.
}%
\label{fig:lens_vel_inv}
\end{figure}

\begin{figure}[ht!]
\begin{center}
\includegraphics[width=0.62\textwidth]{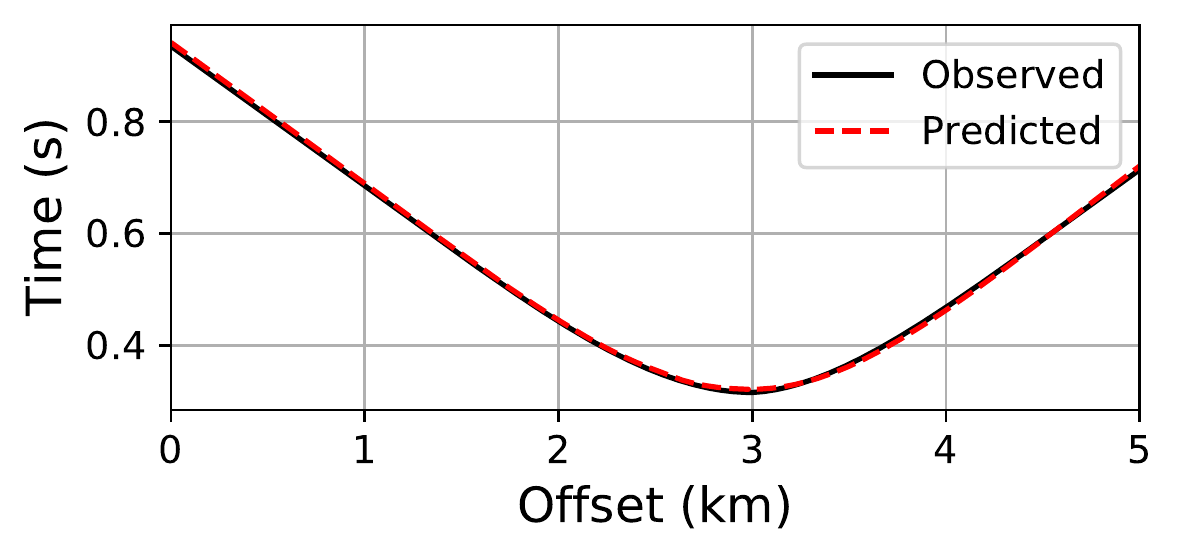}
\end{center}
\caption{
A comparison between the observed traveltimes (solid black) and those predicted using the neural network (dashed red) at the receivers located on the surface for the source at $x~=3$~km.
}%
\label{fig:lens_datafit}
\end{figure}

Figure~\ref{fig:lens_vel_inv} shows the inverted velocity model, indicating that the long-wavelength features, in particular for the velocity bump between $x = 2-3$~km have been well-recovered. We also observe a good fit between the observed and predicted traveltimes at the receivers, as shown in Figure~\ref{fig:lens_datafit}. It is worth mentioning that for conventional tomography, a depth gradient for the initial velocity is necessary, and the choice of it may bear some consequences on the inverted velocities. However, as highlighted earlier, PINNtomo is agnostic to the initial velocity distribution and does not require an initial model with a depth gradient, even for surface tomography.

\section{Discussion and conclusions}

In this work, we presented a novel approach to the traveltime tomography problem. To this end, we use neural networks to approximate the traveltime factor and the velocity fields, subject to the physics-informed regularization based on the eikonal equation. Through tests on cross-hole and surface acquisition geometries, we observe that the method is capable of reliably recovering the long wavelength features of the velocity field. Conventional techniques require an initial depth gradient for velocity that may affect the inversion result. On the contrary, the performance of the proposed approach is largely agnostic to the initial velocity model. 

The proposed approach enjoys several advantages compared to conventional tomography methods. Typically, conventional methods use some form of smoothing regularization to compensate for the poor illumination of the model space due to the acquisition geometry. These physics-agnostic regularizers end up limiting the resolution of the inverted velocity model.  On the contrary, using the residual of the eikonal equation in the loss function, PINNtomo enforces a physics-informed regularizer, which uses the actual wave propagation physics to address the ill-posedness of the problem. Moreover, unlike conventional methods, the proposed approach is mesh-free, and therefore, can be easily used for models with irregular topography and can also accommodate source and receiver points that are not on a regular grid. Furthermore, through the use of transfer learning, the approach is well-suited to study temporal variations in the near-surface using surface seismic tomography or at larger depths using cross-hole tomography.

The method also outperforms conventional techniques in terms of both memory and computational requirements. While the adjoint-state tomography method overcame the memory limitation of the ray method by avoiding the need to explicitly compute the Fr\'{e}chet derivative matrix, it still depends on the size of the velocity model, which could still be cumbersome for large 3D surveys. Here, the memory required is dependent on the optimization batch size, which is usually much lower than the entire model size. Moreover, since PINNtomo uses \texttt{Tensorflow} at the backend, it allows easy deployment of computations across a variety of platforms (CPUs, GPUs) and architectures (desktops, clusters). Therefore, in this study, we used an NVIDIA Tesla P100 GPU that required only $\sim6$ minutes to invert the velocity model for the cross-hole example and $\sim20$ minutes for the surface tomography example.

\bibliographystyle{unsrt}  
\bibliography{references}  


\end{document}